\documentstyle[twoside,fleqn,espcrc2,epsf]{article}
\input{epsf}

\title{Depinning and dynamic phases in driven three-dimensional vortex lattices
in anisotropic superconductors}
\author{C. J. Olson
\address{Department of Physics, University of California, Davis, California
95616}
and
N. Gr{\o}nbech-Jensen
\address{Department of Applied Science, University of California, Davis, 
California 95616}
\address{NERSC, Lawrence Berkeley National Laboratory, Berkeley, 
California 94720}
}

\begin{document}

\begin{abstract}
We use three-dimensional molecular dynamics simulations of magnetically
interacting pancake vortices to study the dynamic phases of vortex lattices
in highly anisotropic materials such as BSCCO.  Our model treats the
magnetic interactions of the pancakes exactly, with long-range logarithmic
interactions both within and between planes.  The pancake vortices decouple
at low drives and show 
two-dimensional plastic flow.  The vortex lattice both recouples
and reorders as the driving current is increased, eventually forming
a recoupled crystalline-like state at high drives.  
We construct a phase diagram as a function of interlayer
coupling and 
show the relationship between the recoupling transition and the 
single-layer reordering transitions.
\vspace{1pc}
\end{abstract}

\maketitle

In highly anisotropic superconductors such as BSCCO,
the vortex lattice is composed of individual pancake vortices that
may be either coupled or decoupled between layers depending on 
such factors as the material stoichiometry or 
the magnitude and angle of the applied magnetic field.
Of particular interest is the possible relationship between
a coupling/decoupling transition and the widely studied second peak or fishtail
effect \cite{blatter}.

As a function of interlayer coupling strength $s$, there are
two limits of vortex behavior in a system containing pointlike disorder.  
For zero interlayer coupling $s=0$, each plane behaves as
an independent two-dimensional (2D) system.  
For infinite interlayer coupling $s=\infty$,
the vortices form perfectly straight three-dimensional (3D) 
lines, and all of the planes move in unison. 
At finite coupling strength $s\ne 0$, a transition
between these types of behaviors should
occur with coupling strength, but it is unclear whether this transition
is sharp or if an intermediate state of the lattice exists.
Furthermore, it is known that 2D systems with pointlike pinning
can exhibit dynamic reordering under the influence of an applied driving
current,  passing from a liquid-like state at zero drive
to a recrystallized state at high current \cite{reorder,olson,kolton}.  
Thus, in a 3D system, 
a dynamically driven recoupling transition
could be expected, but it is unclear where this
transition falls in relation to the 2D reordering transitions already
seen. 

To study the coupling transitions, we have developed a simulation containing
the correct magnetic interactions between pancakes \cite{clem}.
This interaction is long range both in and between planes, and is
treated according to Ref.\ \cite{ngj}. 

The overdamped equation of motion, $T=0$, for vortex $i$ is given by
$ {\bf f}_{i} = \sum_{j=1}^{N_{v}}\nabla {\bf U}(\rho_{i,j},z_{i,j})
+ {\bf f}_{i}^{vp} + {\bf f}_{d}= {\bf v}_{i}$,
where $N_v$ is the number of vortices, $\rho$ and $z$ are the distance
between pancakes in cylindrical coordinates.
The magnetic energy between pancakes is 
\begin{eqnarray}
{\bf U}(\rho_{i,j},0)=2d\epsilon_{0} 
\left((1-\frac{d}{2\lambda})\ln{\frac{R}{\rho}}
+\frac{d}{2\lambda} 
E_{1}(\rho)
\right) 
\nonumber
\end{eqnarray}
\begin{eqnarray}
{\bf U}(\rho_{i,j},z)=-s\frac{d^{2}\epsilon_{0}}{\lambda}
\left(\exp(-z/\lambda)\ln\frac{R}{\rho}- 
E_{1}(R)
\right) \nonumber
\end{eqnarray}
where
$R = \sqrt{z^2 + \rho^2}$, 
$E_{1}(x) = \int^{\infty}_{\rho}\exp(-x/\lambda)/\rho^{\prime}$ and
$\epsilon_{0} = \Phi_{0}^{2}/(4\pi\xi)^{2}$.
The pointlike pins are randomly distributed in each layer and modeled
by parabolic traps.
We vary the relative strength of the interlayer coupling using the
prefactor $s$.
We have simulated a $16\lambda \times 16\lambda$ 
system containing 89 vortices and 4 layers, with
a total of 356 pancake vortices.  Further work on systems containing
up to 16 layers will be reported elsewhere \cite{toappear}.

In Fig.\ 1(a) we present 
a phase diagram as a function of interlayer coupling strength
$s$ and driving force $f_d$.  
At zero drive, we find a recoupling transition for a coupling
strength of $s> 4.5$.  In samples with $s\ge 5$, the pancakes
remain coupled into lines at all drives and show the same
transitions seen in previous work \cite{olson},
exhibiting plastic flow
of stiff lines above depinning, and reordering into
first a smectic state and then a recrystallized state of stiff
lines at higher drives.  

For samples with weaker interlayer coupling, $s<5$, the vortex lattice
is broken into decoupled planes at zero drive. 
Upon application of a driving current, the samples exhibit 2D
plastic flow in which each layer moves independently of the others.
Once the individual layers reach the driving force at which a
transition to a smectic state occurs, the vortices simultaneously
form the smectic state and recouple, as can be seen from the measure
shown in Fig.\ 1(c)
of the $z$-axis correlation $C_z = 1 - \langle({\bf r}_{i,L}-{\bf r}_{i,L+1})
\Theta( a_0/2 - |({\bf r}_{i,L}-{\bf r}_{i,L+1})|)\rangle$,
where 
$a_0$ is the vortex lattice constant.
The dynamic recoupling transition line follows the smectic
transition line down to $s=2$
and is associated with a peak in the $dV/dI$ curve
seen in Fig.\ 1(b).  
Both the static and dynamic transition lines between decoupled 2D and 
recoupled 3D behavior are sharp.

As a function of the number of layers, we observe the same behavior,
but the depinning current
in the 3D stiff state drops with the number of layers. 
The depinning current in the 2D decoupled phase is not affected since
in this case the individual planes behave as isolated entities.
The recoupling transition sharpens as the number of
layers is increased \cite{toappear}.

We acknowledge helpful discussions with L. N. Bulaevskii, A. Kolton,
C. Reichhardt, R.T. Scalettar, G. T. Zim{\' a}nyi. 
This work was supported by CLC and CULAR (LANL/UC) and by
the Director, Office of Adv.\ Scientific
Comp.\ Res., Div.\ of Math., Information, and 
Comp.\ Sciences, U.S.\ DoE contract DE-AC03-76SF00098.

\begin{figure}
\centerline{
\epsfxsize=7cm 
\epsfbox{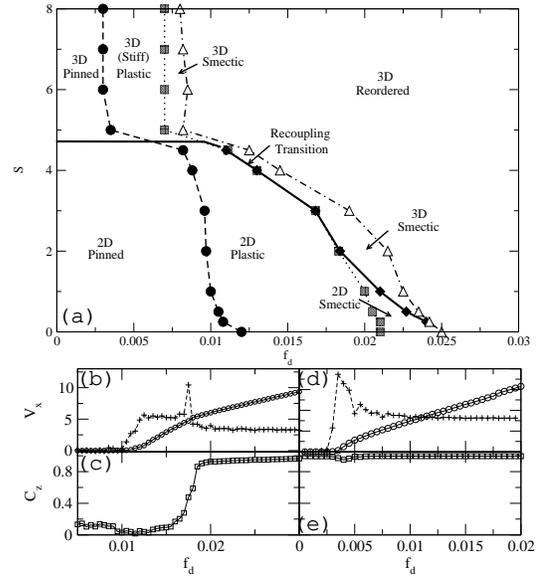}}
\caption{(a) Phase diagram for varying interlayer coupling $s$ and driving
force $f_d$.  Circles: depinning line; diamonds: decoupling line; 
squares: smectic transition line; triangles: recrystallization line.  
(b) $V_x$ (circles) and $dV/dI$ (+ signs) for $s=2.0$.
(c) $C_z$ for $s=2.0$
(d) $V_x$ (circles) and $dV/dI$ (+ signs) for $s=8.0$.
(e) $C_z$ for $s=8.0$}
\label{fig:fig1}
\end{figure}

\end{document}